\documentclass[aps,pra,floatfix,twocolumn,superscriptaddress,showpacs,10pt]{revtex4-1}
\usepackage{amssymb,amsmath,color,mciteplus,graphicx,subfigure}
\usepackage{calc,epsfig,epstopdf,color,mciteplus,bm,mathrsfs}
\usepackage{times}
\usepackage{float}
\usepackage{lipsum}
\usepackage{color}
\usepackage{amsmath}
\usepackage{braket}
\usepackage[unicode=true,
bookmarks=true,bookmarksnumbered=false,bookmarksopen=false,
breaklinks=false,pdfborder={0 0 1},backref=false,colorlinks=true]
{hyperref}
\hypersetup{linkcolor=magenta, urlcolor=blue, citecolor=blue,pdfstartview={FitH}, hyperfootnotes=false, unicode=true}

\begin{document}

\title{Robustness Enhancement of Universal Noncyclic Geometric Gates via Evolution Optimization}
%Robustness Enhancement of Universal Noncyclic Geometric Gates via Geometric Condition Optimization

\author{Zi-Hao Qin} %\email{zihaoqin2004@163.com}
\affiliation{Key Laboratory of Atomic and Subatomic Structure and Quantum Control (Ministry of Education), Guangdong Basic Research Center of Excellence for Structure and Fundamental Interactions of Matter, and School of Physics, South China Normal University, Guangzhou 510006, China}

\author{Yan Liang} \email{liangyan9009@163.com}
\affiliation{School of Physical Science and Technology, Guangxi Normal University, Guilin 541004, China}
%\affiliation{Key Laboratory of Atomic and Subatomic Structure and Quantum Control (Ministry of Education), Guangdong Basic Research Center of Excellence for Structure and Fundamental Interactions of Matter, and School of Physics, South China Normal University, Guangzhou 510006, China}

\author{Yi-Han Yuan} 
\affiliation{Key Laboratory of Atomic and Subatomic Structure and Quantum Control (Ministry of Education), Guangdong Basic Research Center of Excellence for Structure and Fundamental Interactions of Matter, and School of Physics, South China Normal University, Guangzhou 510006, China}

\author{Zheng-Yuan Xue} \email{zyxue83@163.com}
\affiliation{Key Laboratory of Atomic and Subatomic Structure and Quantum Control (Ministry of Education), Guangdong Basic Research Center of Excellence for Structure and Fundamental Interactions of Matter, and School of Physics, South China Normal University, Guangzhou 510006, China}
\affiliation{Guangdong Provincial Key Laboratory of Quantum Engineering and Quantum Materials, Guangdong-Hong Kong Joint Laboratory of Quantum Matter,  and Frontier Research Institute for Physics,\\ South China Normal University, Guangzhou 510006, China}

\author{Tao Chen} \email{chentamail@163.com}
\affiliation{Key Laboratory of Atomic and Subatomic Structure and Quantum Control (Ministry of Education), Guangdong Basic Research Center of Excellence for Structure and Fundamental Interactions of Matter, and School of Physics, South China Normal University, Guangzhou 510006, China}
\affiliation{Guangdong Provincial Key Laboratory of Quantum Engineering and Quantum Materials, Guangdong-Hong Kong Joint Laboratory of Quantum Matter,  and Frontier Research Institute for Physics,\\ South China Normal University, Guangzhou 510006, China}

%\date{\today}

\begin{abstract}
Noncyclic geometric gates aim to overcome the stringent constraints of conventional cyclic conditions and enhance the flexibility in evolution choice. Conceptually, they can also avoid the error problems arising from the violation of cyclicity, thus holding significance for improving the fault tolerance of quantum gates. However, current research on noncyclic geometric gates lacks a comprehensive exploration of their flexibility in evolution choice and validation of their effectiveness in resilience against multiple error sources present in practical quantum systems. In this paper, we systematically evaluate all noncyclic evolution conditions, elucidate their corresponding potential geometric trajectories, and identify optimal conditions for enhancing gate robustness by quantifying the resilience of constructed noncyclic geometric gates against universal systematic errors and residual crosstalk error. The optimized gates demonstrate significant robustness advantages over representative dynamical Rabi gates and conventional cyclic geometric gates. Furthermore, we validate the physical feasibility of high-fidelity noncyclic geometric gates in superconducting quantum circuits, with focused investigation into the impacts of intrinsic leakage errors and decoherence effects. Therefore, this work establishes a critical foundation for robust gate implementation toward practical fault-tolerant quantum processors.
\end{abstract}
\maketitle

\section{Introduction}

The core of quantum computing lies in the precise manipulation of quantum bits (qubits), while the fault tolerance of quantum gates constitutes a key bottleneck for achieving practical quantum computation \cite{QCQI}. Geometric quantum computation \cite{NGQC+NHQC, NHQC+OC} utilizes the geometric phases of quantum states to implement quantum gate operations. Its fundamental advantage stems from the geometric phases \cite{BerryP, Ad-nonAb, AAphase, noncyclicP, nonAd-nonAbe} depending solely on the overall geometric path of the system's evolution, thereby exhibiting inherent robustness against specific noise in the control parameters during the evolution \cite{robust1, robust2, robust3, robust4}. Notably, nonadiabatic geometric quantum gates \cite{NGPG, NGQC, NHQC, NHQC+DFS, SL-NHQC, SL2-NGQC, OP+NHQC, TC1-NGQC, GSC-NGQC, TC2-NGQC}, which overcome the limitations of the adiabatic approximation, have been experimentally realized on diverse quantum platforms due to their advantages of fast and simplified control, including trapped ion \cite{Geo-ion1, Geo-ion2}, nuclear magnetic resonance \cite{Geo-NMR1, Geo-NMR2, Geo-NMR3, Geo-NMR4, Geo-NMR5}, superconducting quantum circuits \cite{Geo-SQ1, Geo-SQ2, Geo-SQ3, Geo-SQ4, Geo-SQ5, Geo-SQ6, Geo-SQ7}, nitrogen-vacancy centers \cite{Geo-NV1, Geo-NV2, Geo-NV3, Geo-NV4, Geo-NV5, Geo-NV6}, etc. However, conventional geometric quantum gates usually strictly adhere to the cyclic evolution condition to ensure the precise accumulation of geometric phases. This constraint of cyclicity significantly restricts the flexibility in evolution choice and may introduce additional errors under non-ideal experimental environments.

The concept of noncyclic geometric quantum gates thus emerged \cite{Noncyc-NGQC1, Noncyc-NGQC2, Noncyc-NGQC3, Noncyc-NGQC4, Noncyc-NGQC5, Noncyc-NGQC6}, aiming to break through the strict limitations of the conventional cyclic evolution condition, thereby providing greater freedom in evolution choice. In principle, this flexibility not only facilitates the optimization of gate operation performance but also circumvents errors arising from the violation of the cyclic evolution condition, demonstrating significant potential for enhancing the fault tolerance of quantum gates. High-fidelity nonadiabatic geometric gates have already been experimentally validated in both atomic systems \cite{NNGeo-atom} and superconducting quantum circuits \cite{NNGeo-SQ}. Current research on noncyclic geometric gates still exhibits notable deficiencies: on the one hand, there is a lack of systematic exploration of the flexibility in evolution choice; on the other hand, their effectiveness in resisting multiple error sources has not yet been fully verified.
%(such as systematic control errors, residual crosstalk, etc.)

Here, we will focus on addressing the aforementioned key issues. To this end, we systematically evaluate all possible noncyclic evolution conditions and provide an in-depth elucidation of their corresponding potential geometric trajectories. By quantifying the robustness of the constructed noncyclic geometric gates against universal systematic errors and residual crosstalk error, we identify the optimal evolution conditions capable of enhancing gate robustness. The optimized noncyclic geometric gates demonstrate significant robustness advantages over representative dynamical Rabi gates and conventional cyclic geometric gates. Furthermore, we validate the physical feasibility of high-fidelity noncyclic geometric gates in superconducting quantum circuits, with focused investigation into the impacts of intrinsic leakage errors and decoherence effects. Collectively, this work offers an attractive avenue for realizing practical quantum computation.

\section{Noncyclic Geometric Evolution: A Comprehensive Analysis}

We begin by considering a general two-level quantum system comprising a ground state $|0\rangle$ and an excited state $|1\rangle$. Arbitrary quantum control of this system can be achieved through external microwave field driving. Under the rotating wave approximation, the Hamiltonian in the interaction picture can be expressed as:
\begin{eqnarray} \label{GeneralH}
\mathcal{H}(t)=\frac {1} {2}
\left(
\begin{array}{cccc}
 -\Delta(t)             & \Omega(t) e^{-i\phi(t)} \\
 \Omega(t) e^{i\phi(t)} & \Delta(t)
\end{array}
\right),
\end{eqnarray}
where $\Omega(t)$ and $\phi(t)$ are the time-dependent driving amplitude and phase of the microwave field, respectively; $\Delta(t)$ is the detuning with respect to the frequency difference between the microwave field and the two-level system.

\begin{figure}[t]
  \centering
\includegraphics[width=1.0\linewidth]{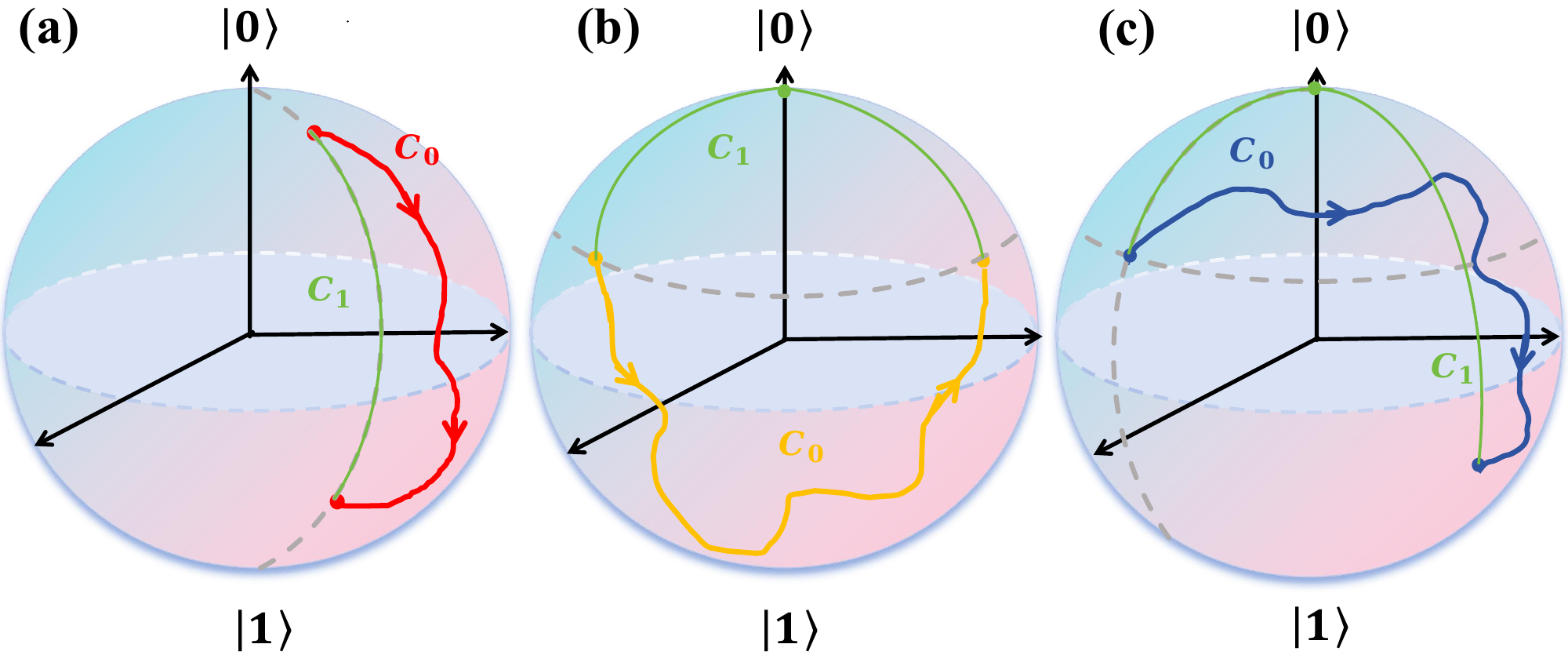}  
\caption{The potential noncyclic geometric evolution trajectories corresponding to (a) noncyclic evolution condition (i), (b) condition (ii), and (c) condition (iii), denoted as $C_0$. The geodesic $C_1$, which connects the start and end points of the actual noncyclic geometric trajectory $C_0$, is represented by a green solid line.}
\label{Figure1}
\end{figure}

To accurately describe the evolution operator associated with the aforementioned Hamiltonian, the following two orthogonal state vectors are selected for analysis:
\begin{subequations} \label{TwoState}
\begin{align}
|\Psi_1(t)\rangle&=e^{if_1(t)}[\cos\frac{\chi(t)}{2}|0\rangle+\sin\frac{\chi(t)}{2} e^{i\xi(t)}|1\rangle], \\
|\Psi_2(t)\rangle&=e^{if_2(t)}[\sin\frac{\chi(t)}{2}e^{-i\xi(t)}|0\rangle -\cos\frac{\chi(t)}{2} |1\rangle],
\end{align}
\end{subequations}
where $\chi(t)$ and $\xi(t)$ are the spherical coordinates of the state vector on the Bloch sphere as shown in Fig. \ref{Figure1}. $f_{k}(t)$ represents the overall phase accumulated, with the initial value $f_{k}(0)=0$. The evolution detail of the state vector $|\Psi_k(t)\rangle$ is governed by the Hamiltonian $\mathcal{H}(t)$, and they are connected by the Schr\"{o}dinger equation $i(\partial/\partial t) |\Psi_k(t)\rangle =\mathcal{H}(t)|\Psi_k(t)\rangle$. Therefore, by characterizing the target evolution process governed by $\chi(t)$ and $\xi(t)$, the Hamiltonian control parameters $\{\Delta(t), \Omega(t), \phi(t)\}$ can be reverse engineered accordingly (for detail, see Appendix A). In this way, the correspondence between parameters can be expressed as $\dot{\chi}(t)=\Omega(t)\sin[\phi(t)-\xi(t)]$ and $\dot{\xi}(t)=-\Delta(t)-\Omega(t)\cot\chi(t)\cos[\phi(t)-\xi(t)]$. In addition, we can further calculate the total phase accumulated during the evolution period $\tau$ as $\gamma=f_1(\tau)=-f_2(\tau)=\frac{1}{2}\int^{\tau}_0 \{\dot{\xi}(t)\left[1-\cos\chi(t)\right]+\Delta(t)\}/\cos\chi(t) \textrm{d}t$.

%\red{The geometric phase component in $\gamma$ can be formally expressed as 
%\begin{equation} \label{GammaG}
%\gamma_g=\gamma-\gamma_d=-\frac {1} {2}\int^\tau_0 \dot{\xi}(t)\left[1-\cos\chi(t)\right] \textrm{d}t,
%\end{equation}
%where $\gamma_d=-\int^\tau_0\langle \Psi_1(t)|\mathcal{H}(t)|\Psi_1(t)\rangle \textrm{d}t$ corresponds to the dynamical phase that is directly associated with the Hamiltonian parameters. The geometric nature \red{[99,99]} of $\gamma_g$ comes from the fact that it is given by half of the solid angle enclosed by the noncyclic evolution path and its geodesic connecting the initial point $[\chi(0),\xi(0)]$ to the final point $[\chi(\tau),\xi(\tau)]$. Therefore, by ensuring that $\gamma_d=0$, the evolution of this system can be a purely geometric process.}

Taking the state vector $\ket{\Psi_{1}(t)}$ as the illustration, when it undergoes noncyclic evolution—meaning its evolutionary trajectory is not closed—the total phase difference $\gamma$ between the initial state $\ket{\Psi_{1}(0)}$ and the final state $\ket{\Psi_{1}(\tau)}$ can also be expressed as $\gamma=\text{Arg}\braket{\Psi_{1}(0)|\Psi_{1}(\tau)}$. After eliminating the dynamical phase component $\gamma_d=-\int^\tau_0\langle \Psi_1(t)|\mathcal{H}(t)|\Psi_1(t)\rangle \textrm{d}t$, the geometric phase \cite{noncyclicP, PPhase2} accumulated along the noncyclic trajectory $C_0$ is obtained:
\begin{equation} \label{GammaG}
\gamma_p(C_0)=\text{Arg}\braket{\Psi_{1}(0)|\Psi_{1}(\tau)}+\int^\tau_0\langle \Psi_1(t)|\mathcal{H}(t)|\Psi_1(t)\rangle \textrm{d}t.
\end{equation} 
For such noncyclic geometric evolution, assume that there exists a geodesic $C_1$ connecting the initial point $[\chi(0),\xi(0)]$ and the final point $[\chi(\tau),\xi(\tau)]$ of the actual evolution trajectory $C_0$, such that $C_0$ and $C_1$ together form a closed trajectory $C$, as shown in Fig. \ref{Figure1}. The total phase difference along this closed trajectory $C$ (whose geometric element is the A-A phase \cite{AAphase}) can be expressed as $\gamma_p(C) = \gamma_p(C_0) + \gamma_p(C_1) = \frac{1}{2}\int_C \sin \chi \textrm{d}\chi \textrm{d}\xi$, which corresponds to half of the solid angle enclosed by the closed trajectory $C$. Following the geodesic rule \cite{noncyclicP, PPhase2}, the additional geodesic $C_1$ contributes no phase accumulation since $\textrm{d}\xi=0$. Thus, the geometric phase accumulated along the noncyclic trajectory $C_0$ is regarded as being identical to the cyclic geometric phase corresponding to the closed trajectory, namely, $\gamma_g=-\frac {1} {2}\int^\tau_0 \dot{\xi}(t)\left[1-\cos\chi(t)\right] \textrm{d}t$, which also has a gauge-invariant and geometric nature.
%几何非循环演化过程的优势

Upon elimination of the dynamical phase during the evolution process (i.e., enforcing $\gamma_d=0$), the two orthogonal state vectors in Eq. (\ref{TwoState}) undergo a complete geometric evolution from $|\Psi_{1,2}(0)\rangle$ to $|\Psi_{1,2}(\tau)\rangle$ by the final evolution time. The geometric evolution operator can then be formulated as:
\begin{equation} \label{generalU}
%\begin{small}
\begin{aligned}
&U(\tau)
=\sum_{k=1,2}|\Psi_{k}(\tau)\rangle \langle\Psi_{k}(0)| \\
&= \left[\begin{array}{cc}
e^{-i\xi_-}(M_{\gamma^\prime,\chi_-}\!+ iQ_{\chi_+,\gamma^\prime}
) 
& \!e^{-i\xi_+}(iP_{\gamma^\prime,\chi_+}
\!-Q_{\gamma^\prime,\chi_-}) 
\\
e^{-i\xi_+}(iP_{\gamma^\prime,\chi_+}
\!+Q_{\gamma^\prime,\chi_-}) 
& \!e^{-i\xi_-}(M_{\gamma^\prime,\chi_-}
\!-iQ_{\chi_+,\gamma^\prime})
\end{array}\right], \\
\end{aligned}
%\end{small}
\end{equation}
in which $P_{a,b}=\sin a\sin b$, $Q_{a,b}=\cos a\sin b$, $M_{a,b}=\cos a\cos b$, $\xi_\pm=[\xi(\tau)\pm\xi(0)]/2$, $\chi_\pm=[\chi(\tau)\pm\chi(0)]/2$, and $\gamma'=\gamma+\int_{0}^{\tau}\dot{\xi}(t)/2$. The specific type of the resulting noncyclic geometric gate is determined by the boundary values of $\chi(t)$ and $\xi(t)$.

The utilization of noncyclic geometric phases allows the final system evolution to selectively fulfill one of three noncyclic evolution conditions:
\begin{subequations}
\begin{align}
\text{(i)} \quad \chi(\tau)\neq\chi(0) \quad &\text{and} \quad  \xi(\tau)=\xi(0)\pm2n\pi, \\
\text{(ii)} \quad \chi(\tau)=\chi(0) \quad &\text{and} \quad \xi(\tau)\neq\xi(0)\pm2n\pi, \\
\text{(iii)} \quad \chi(\tau)\neq\chi(0) \quad &\text{and} \quad \xi(\tau)\neq\xi(0)\pm2n\pi,
\end{align}
\end{subequations}
where $n$ denotes the total number of rotations corresponding to the system's evolution along the latitudinal lines. As shown in Fig. \ref{Figure1}, different noncyclic evolution conditions correspond to different actual geometric trajectories. Consequently, these three conditions provide enhanced flexibility in evolution choice while circumventing the stringent requirement of satisfying the single cyclic evolution condition, $\chi(\tau)=\chi(0)$ and $\xi(\tau)=\xi(0)\pm2n\pi$, inherent in cyclic geometric evolution. Moving forward, we aim to determine from these the optimal noncyclic evolution conditions capable of effectively improving gate robustness, thereby facilitating the realization of high-fidelity noncyclic geometric quantum gates.

\section{Design framework for universal noncyclic geometric gates}

Next, for intuitive comparison, when analyzing the specific geometric evolution trajectories corresponding to each noncyclic evolution condition, we only consider trajectory segments along longitudinal and latitudinal lines. In addition, we will address key errors in quantum systems—including universal systematic errors and residual crosstalk error—to identify the optimal noncyclic evolution condition that effectively mitigate their impacts.

\subsection{Noncyclic Evolution Condition (i)}

\begin{figure}[t]
  \centering
  \includegraphics[width=1.0\linewidth]{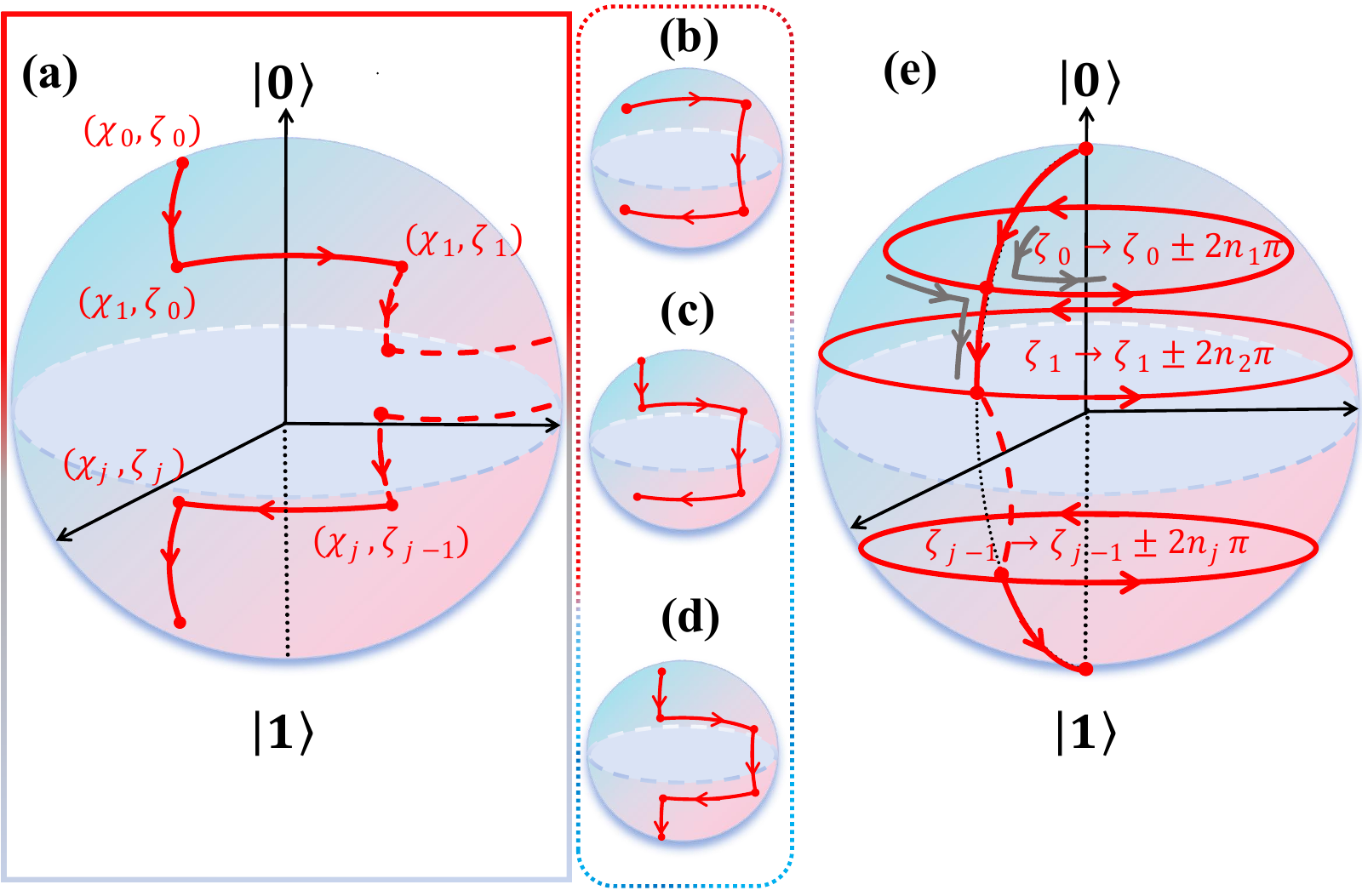}
  \caption{(a) The general geometric evolution trajectory corresponding to noncyclic evolution condition (i) when $n=0$, where only trajectory segments along longitudinal and latitudinal lines are considered. (b)-(d) Several simplified evolution trajectories based on noncyclic evolution condition (i). (e) The general geometric evolution trajectory corresponding to noncyclic evolution condition (i) when $n\neq0$, featuring intermediate trajectory segments with multiple loops.}
\label{Figure2}
\end{figure}

As shown in Fig. \ref{Figure1}(a), for the noncyclic evolution condition (i), the polar angles of the initial and final positions of the potential geometric evolution trajectory are different (that is, $\chi(\tau)\neq\chi(0)$), but the azimuth angles of the initial and final positions will still remain on the same meridian after the trajectory evolves along the latitudinal lines for $n$ circles (that is, $\xi(\tau)=\xi(0)\pm2n\pi$). In this case, the noncyclic geometric evolution operator derived from noncyclic evolution condition (i) can be further simplified by Eq. (\ref{generalU}) to the following form:
\begin{equation} \label{U_C1}
\begin{small}
U(\tau)
\!=\! \left[\begin{array}{cc}
iQ_{\chi_+,\gamma^\prime}
\!+\!M_{\gamma^\prime,\chi_-} & \!e^{-i\xi_0}(iP_{\gamma^\prime,\chi_+}
\!-\!Q_{\gamma^\prime,\chi_-}) \\
\!e^{-i\xi_0}(iP_{\gamma^\prime,\chi_+}
\!+\!Q_{\gamma^\prime,\chi_-}) & M_{\gamma^\prime,\chi_-}
\!-\!iQ_{\chi_+,\gamma^\prime,}
\end{array}\right],
\end{small}
\end{equation}
in which the parameters of the starting point of the trajectory, $(\chi(0), \xi(0))$, are abbreviated as $(\chi_0, \xi_0)$. Therefore, by setting $\chi(\tau)+\chi(0)=\pi/2$, $\gamma' = \pi/2$, and $\xi_0 = 0$, we can ultimately construct the geometric Hadamard gate $H^i$ based on the noncyclic evolution condition (i). Similarly, for geometric rotation operations around the $X$-axis ($Y$-axis), denoted as $R^i_{x}(\theta)$ ($R^i_{y}(\theta)$), these gates are obtained by imposing $\chi(\tau)-\chi(0)=\theta$, $\gamma'=\pi$, and $\xi_0=-\pi/2$ ($\xi_0=0$ for $R^i_{y}(\theta)$).

In the following, we present the concrete noncyclic geometric evolution trajectories required to implement the target noncyclic geometric gates based on the noncyclic evolution condition (i). We first analyze the case of $n=0$, corresponding to a general noncyclic geometric evolution trajectory as shown in Fig. \ref{Figure2}(a). For brevity, we here uniformly adopt $(\chi_j, \xi_{j-1})$ and $(\chi_j, \xi_j)$ to denote the starting-point and endpoint parameters of the $j$-th latitudinal segment trajectory, respectively. This evolution trajectory starts from the initial point $(\chi_0, \xi_0)$, traverses multiple longitudinal and latitudinal lines, and finally evolves to the endpoint $(\chi_{j+1}, \xi_j)$. To strictly eliminate the dynamical phase, the detuning parameter $\Delta(t)$ must satisfy $\Delta(t)=-\dot{\xi}(t)\sin\chi(t)$. Consequently, in the gate construction, in addition to fixing $[\chi(\tau) \pm \chi(0)]$ and $\xi_0$, this trajectory also requires that the gate parameter
\begin{equation}
\begin{aligned}
\gamma'&=\frac{1}{2}(\xi_1-\xi_0)\cos\chi_1+\frac{1}{2}(\xi_2-\xi_1)\cos\chi_2+\cdots \\
&+ \frac{1}{2}(\xi_j-\xi_{j-1})\cos\chi_j=\frac{1}{2}\sum_{i=1}^{j}(\xi_i-\xi_{i-1})\cos\chi_i.
\end{aligned}
\end{equation}
It can be found that, unlike cyclic geometric schemes, the implementation of noncyclic geometric gates here only requires fixing $[\chi(\tau)\pm\chi(0)]$, without the need to specify the exact value of the evolution parameters $\chi(t)$ at the initial and final moments. This also brings great flexibility to the evolution optimization in the next step.

Clearly, when reducing to the simple trajectories shown in Figs. \ref{Figure2}(b) and \ref{Figure2}(c), it is impossible to simultaneously satisfy $\chi(\tau)-\chi(0)=\pi$ and accumulate a geometric phase at the final moment, thereby preventing direct implementation of the noncyclic geometric $R^i_{x}(\pi)$ and $R^i_{y}(\pi)$ gates via such trajectories. In contrast, the trajectory shown in Fig. \ref{Figure2}(d) can avoid the aforementioned conflict. The Hamiltonian parameters corresponding to the trajectory details can be derived from Appendix A.
Based on this, while ensuring the implementation of the noncyclic geometric $H^i$ gate, we can further optimize the intermediate trajectory parameters $\chi_1$ and $\chi_2$ within the range of $(\chi_0, \pi/2-\chi_0)$ by selecting the initial parameter $\chi_0$. Similarly, for the noncyclic geometric $R^i_{x}(\theta)$ and $R^i_{y}(\theta)$ gates, the intermediate trajectory parameters $\chi_1$ and $\chi_2$ can also be optimized within the range of $(\chi_0, \theta+\chi_0)$ based on the chosen initial parameter $\chi_0$. Obviously, the flexibility in choosing the starting point of noncyclic geometric trajectories also introduces varied optimization designs for intermediate parameters. Through this, the entire noncyclic geometric evolution process is enhanced to achieve strengthened robustness of quantum gates.

Furthermore, when considering the case where $n\neq0$, the trajectory corresponding to that shown in Fig. \ref{Figure2}(e) requires the gate parameter $\gamma'$ to be rewritten as:
\begin{equation}
\gamma'=n_1\pi\cos\chi_1+\cdots+n_j\pi \cos\chi_j=\pi\sum_{i=1}^{j}n_i\cos\chi_i.
\end{equation}
Currently, the optimization of intermediate parameters is replaced by multi-loop (specifically, $n_j$-loop) evolution along different latitudes $\chi_j$. However, this approach increases the evolution time due to its segmented multi-loop trajectories and offers limited optimization for the evolution process, as detailed in Appendix B. Thus, in the discussion of noncyclic evolution condition (i), we identify the noncyclic geometric trajectory at $n=0$ as the optimal trajectory selection under this condition.

\subsection{Noncyclic Evolution Condition (ii)}

Next, for the noncyclic evolution condition (ii), the azimuth angles of the initial and final positions of the potential geometric evolution trajectory, as shown in Fig. \ref{Figure1}(b), are different (that is, $\xi(\tau)\neq\xi(0)\pm2n\pi$), but the polar angles still remain on the same latitudinal line (that is, $\chi(\tau)=\chi(0)=\chi$). Under this condition, the noncyclic geometric evolution operator can be further simplified by Eq. (\ref{generalU}) to the following form:
\begin{equation}
\begin{small}
U(\tau)=\left(\begin{matrix}
e^{i\xi_-}(C_{\gamma'}+iS_{\gamma'}C_{\chi}) 
& e^{-i\xi_+}\ iS_{\gamma'}C_{\chi} 
\\
e^{-i\xi_+}\ iS_{\gamma'}S_{\chi}
& e^{-i\xi_-}(C_{\gamma'}-iS_{\gamma'}C_{\chi})
\end{matrix}\right),
\end{small}
\end{equation}
where $S_a=\sin a$ and $C_a=\cos a$. Thus, by setting $\chi=\pi/2$, $\gamma' = \pi/4$, $\xi_- = \pi$, and $\xi_0=\pi/2$, we can construct the geometric Hadamard gate $H^{ii}$ based on the noncyclic evolution condition (ii). In addition, for geometric rotation operations around the $X$-axis ($Y$-axis), denoted as $R^{ii}_{x}(\theta)$ ($R^{ii}_{y}(\theta)$), these gates can also be obtained by imposing $\chi=\theta/2$, $\gamma'=\pi/2$, $\xi_- = \pi$, and $\xi_0=\pi/2$ ($\xi_0=\pi$ for $R^{ii}_{y}(\theta)$).

\begin{figure}[t]
  \centering \includegraphics[width=1.0\linewidth]{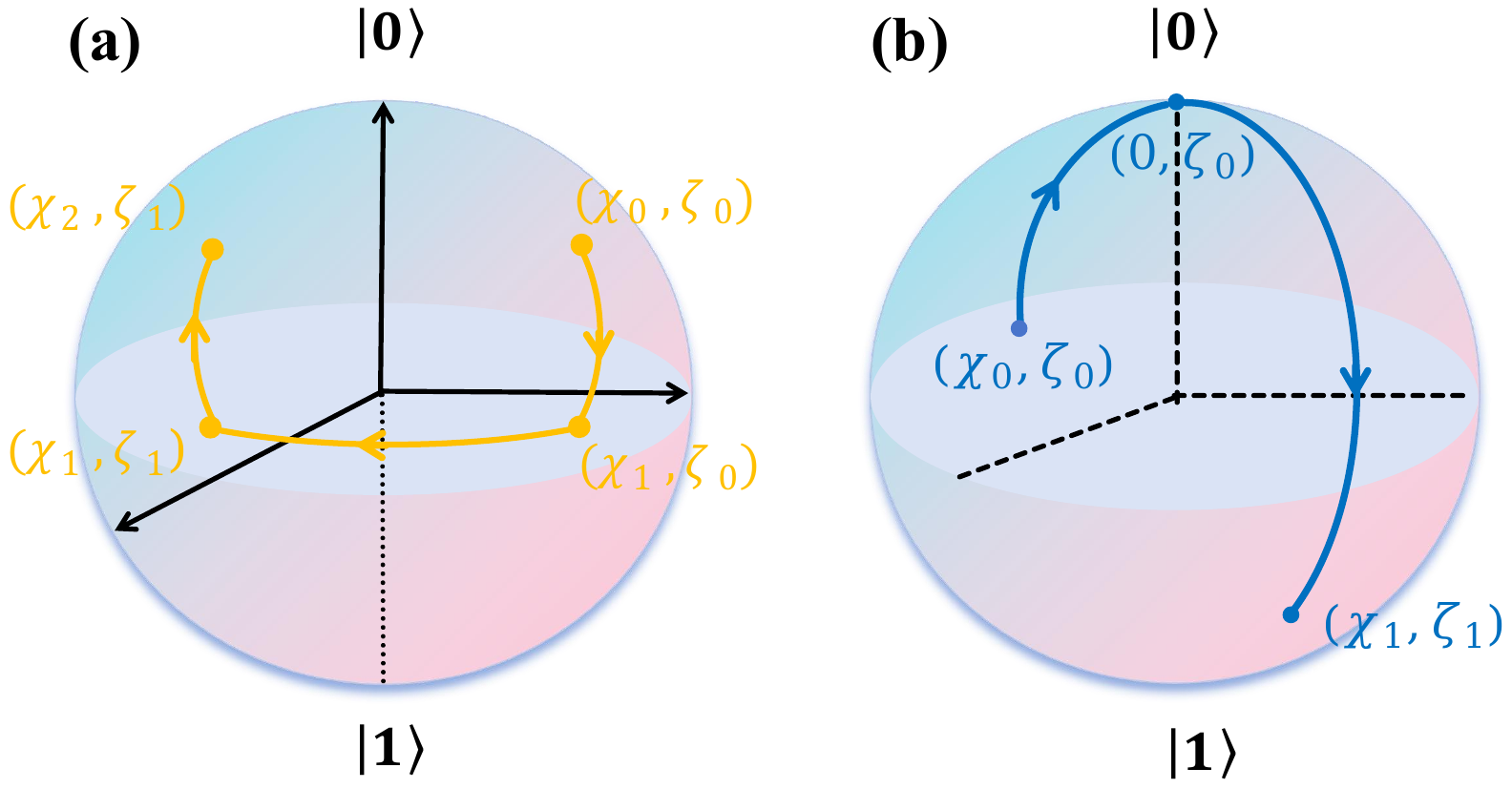}
 \caption{The simplified geometric trajectories corresponding to (a) noncyclic evolution condition (ii) and (b) condition (iii).}
\label{Figure3}
\end{figure}

Similarly, when considering only trajectory segments along longitudinal or latitudinal lines, the simplified evolution trajectory satisfying noncyclic evolution condition (ii) is shown in Fig. \ref{Figure3}(a), with the evolution details as follows:
\begin{align}
&\!\chi(0)=\chi_0 \!\leadsto \chi(\tau_1)=\chi_1 \rightarrow \chi(\tau_2)=\chi_1 \rightarrow \chi(\tau)=\chi_2 \notag\\
&\xi(0)=\xi_0 \rightarrow  \ \ \! \xi(\tau_1)=\xi_0 \leadsto \ \ \! \xi(\tau_2)=\xi_1 \rightarrow \ \ \!\! \xi(\tau)=\xi_1
\end{align}
where the arrows ``$\rightarrow$" and ``$\rightsquigarrow$" indicate whether the parameters $\chi(t)$ and $\xi(t)$ remain constant or change over time during evolution, respectively. This evolution trajectory initiates from the starting point $(\chi_0, \xi_0)$, evolves along a longitudinal line to the point $(\chi_1, \xi_0)$. It then continues along a latitudinal line to the point $(\chi_1, \xi_1)$, and finally evolves along a longitudinal line to the endpoint $(\chi_2, \xi_1)$.

During the evolution along this trajectory, no phase accumulates on the longitudinal lines. We only need to limit the trajectory parameters for the intermediate latitudinal segment to satisfy $-\dot{\xi}(t)\sin\chi_1=\Delta(t)$ to ensure the elimination of the dynamical phase over the entire evolution. Based on the parameter details of the aforementioned noncyclic geometric evolution trajectory and the geometric constraint conditions, the specific requirements for the Hamiltonian parameter settings can be derived from Appendix A. Unlike the geometric trajectory corresponding to noncyclic evolution condition (i), which can be further flexibly designed, the start point $(\chi_0=\chi, \xi_0)$ and end point $(\chi_2=\chi, \xi_1)$ of the trajectory here need to be fixed to construct different quantum gates. Consequently, no additional optimization can be applied to improve the gate performance. Therefore, we select the simple geometric trajectory shown in Fig. \ref{Figure3}(a) as the scheme based on noncyclic evolution condition (ii). This trajectory corresponds to the evolution approach adopted in Ref. \cite{NNGeo-SQ}.

\subsection{Noncyclic Evolution Condition (iii)}

Finally, we focus on the noncyclic evolution condition (iii). Both the azimuthal and polar angles of the initial and final positions of the potential geometric evolution trajectories are not on the same latitude and longitude lines (that is, $\chi(\tau)\neq\chi(0), \xi(\tau)\neq\xi(0)\pm2n\pi$), as shown in Fig. \ref{Figure1}(c). Under this condition, the noncyclic geometric evolution operator remains consistent with Eq. (\ref{generalU}).

Similarly, based on the noncyclic evolution condition (iii), we can realize the noncyclic geometric Hardmard gate $H^{iii}$, by setting $\gamma'=\pi/4, \chi_{+}=-3\pi/2, \chi_{-}=\pi/2, \xi_{\pm} = \mp \frac{\pi}{4}$. In addition, for geometric rotation operations around the $X$-axis and $Y$-axis, denoted as $R^{iii}_{x}(\theta)$ and $R^{iii}_{y}(\theta)$, can also be obtained by imposing $\gamma'=\pi, \xi_- =0, \xi_+=-\pi/2$ for $R^{iii}_{x}(\theta)$; and $\gamma'=\pi, \xi_- =0, \xi_+=\pi$ for $R^{iii}_{y}(\theta)$, where $\theta=\chi_{-}$. 
%and $\gamma'=\pi, \chi_-=0$ for $R^{ii}_{z}(\theta)$.
In fact, any single-qubit $SU(2)$ operation can be realized by choosing phase $\gamma'$, the initial and final value $\xi_{\pm}$ and $\chi_{\pm}$.

Again we only consider the design of trajectories along the latitude and longitude lines, as shown in Fig. \ref{Figure3}(b), with the evolution parameters details $[\xi(\tau)$,$\chi(\tau)]$ as:
\begin{align}
&\!\chi(0)=\chi_0 \leadsto \ \chi\left(\tau_1\right)=\ 0 \ \leadsto\ \chi\left(\tau\right)=\chi_1 \notag\\
&\xi(0)=\xi_0\ \rightarrow  \ \ \! \xi\left(\tau_1\right)=\xi_0\ \rightarrow\ \xi\left(\tau\right)=\xi_1
\end{align}
where the arrows ``$\rightarrow$" and ``$\rightsquigarrow$" also indicate whether the parameters $\chi(t)$ and $\xi(t)$ remain constant or change over time during evolution, respectively. The entire evolution trajectory is divided into two distinct segments, with the initial and end points connected by a geodesic. In the first segment of this evolution shown in Fig. \ref{Figure3}(b), evolutionary state starts from the starting point $(\chi_0, \xi_0)$ and evolves along a longitudinal line to the North pole $(0, \xi_0)$. In the second segment, it then descends to the endpoint $(\chi_1, \xi_1)$ along another longitudinal line.

Due to the purely geometric nature of the evolution process, we need to limit the dynamical phase element $\gamma_d=-\int_{0}^{\tau}{\Omega(t)\text{cos}\chi(t)\text{cos}(\phi(t)-\xi(t))dt}=0$, i.e., $\phi(t)=\xi(t)+\pi/2$ . Based on the parameter details of the noncyclic geometric evolution trajectory and the geometric constraint conditions above, the specific requirements for the Hamiltonian parameter settings can also be derived from Appendix A. Thus, the evolution coordinates $\chi(\tau_1)=0$, $\xi(\tau)=\xi'+\xi(\tau_1)$, where $\xi(\tau_1)\in[0, \pi/2]$, $\xi_0=\xi'+\pi$ ($\xi'$ is a constant), are set to construct the universal quantum gate. According to the evolutionary trajectory above, the universal quantum gate form is determined by four parameters $\left\{\chi_0, \xi_0, \xi_1, \chi_1\right\}$. %Notably, the operator $ U(\tau)(\chi_0, \xi_0, \xi_1, \chi_2) $ is capable of facilitating arbitrary rotations on the Bloch sphere.
For instance, by setting $\xi_0=\pi/2,\xi_1=\pi,\chi_0$ and $\chi_1$ to any appropriate constants, the $R_{x}^{iii}(\theta)$ can be constructed. Similarly, we can realize $R_{y}^{iii}(\theta)$ and $H^{iii}$ gates by setting the following parameters $\left\{\chi_0, -\pi, \pi, \chi_1\right\}$, $\left\{\pi/4, \pi, 0, \pi/4\right\}$ respectively.

Furthermore, any rotation angle $\theta$ for both cases is given by $\chi_1 - \chi_0$. Although there are multiple choices for $\chi_0$ and $\chi_1$ when designing rotations around the $X$-axis or $Y$-axis, the conditions $\chi_0$ and $\chi_1$ must be positive. These constraints ensure that the evolution time remains positive and prevent the gate from degenerating into a trivial pulse sequence. But it also means, at the same time, the absence of additional tunable parameters consequently restricts our ability to further optimize gate fidelity within the framework of condition (iii). This trajectory also corresponds to the evolution approach adopted in Ref. \cite{Noncyc-NGQC3, NNGeo-atom}.

\begin{figure}[htbp]
  \centering \includegraphics[width=1.0\linewidth]{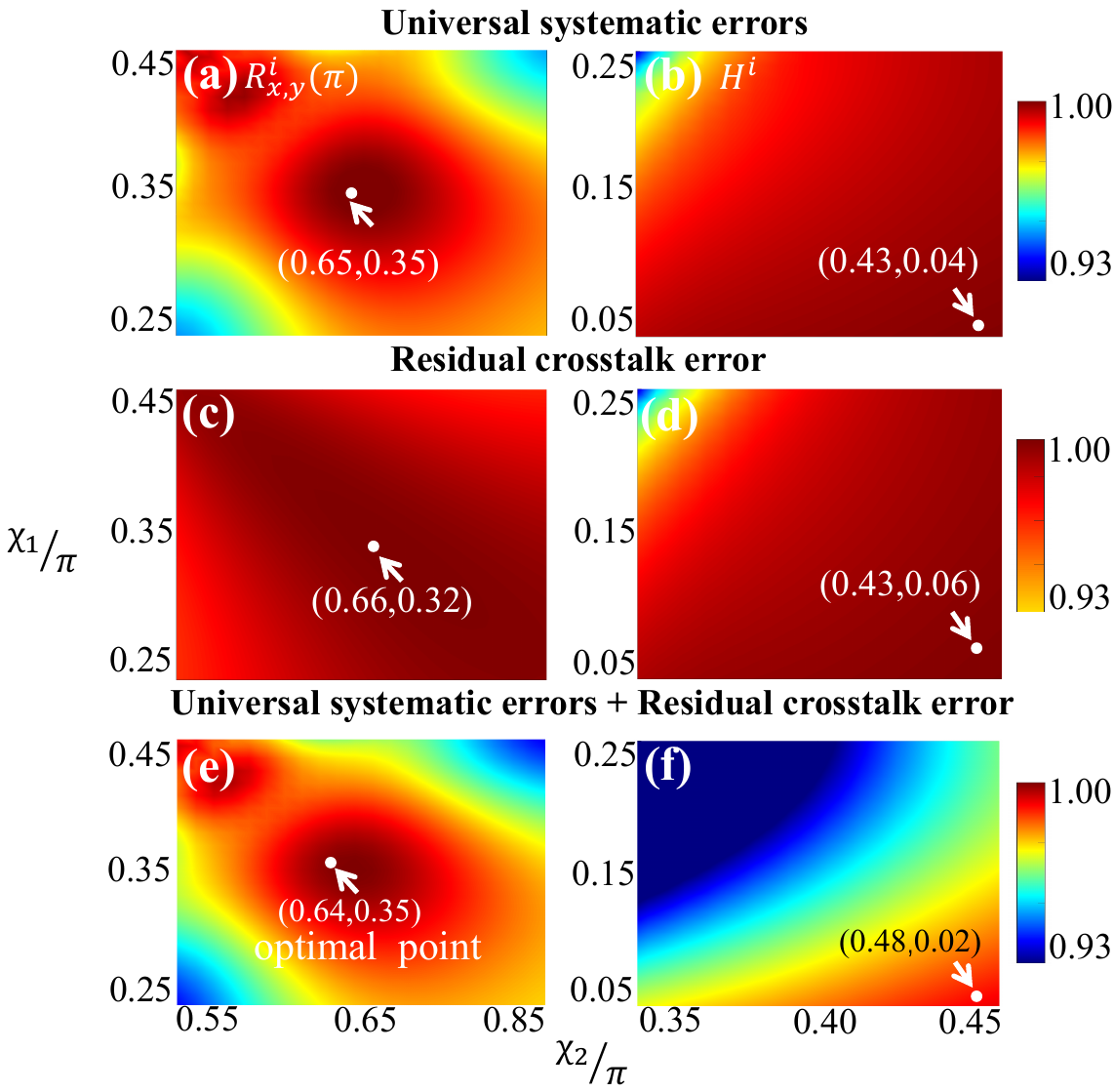}
   \caption{Fidelity of the noncyclic geometric $R^i_{x,y}(\pi)$ ($H^i$) gate as a function of intermediate trajectory parameters $\chi_1$ and $\chi_2$ under the impacts of (a) ((c)) universal systematic errors and (b) ((d)) residual crosstalk error, and the combined impacts of universal systematic errors and residual crosstalk error in (e) (f), with error strength $\lambda_m/\Omega_m=\zeta_{zz}/\Omega_m=0.05$.}
\label{Figure4}
\end{figure}

\begin{figure}[t]
  \centering
  \includegraphics[width=1.02\linewidth]{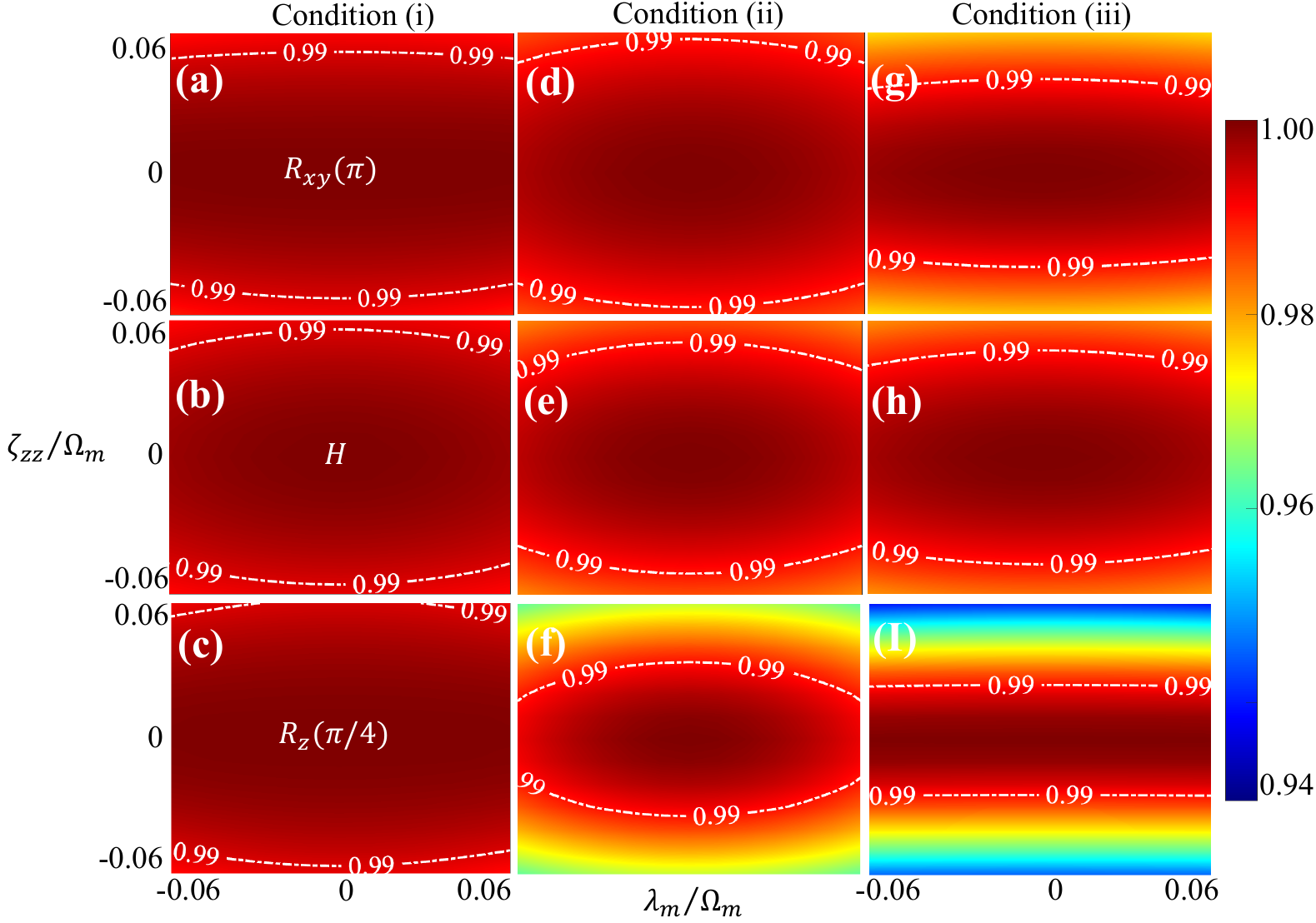}
  \caption{Comparison of gate robustness for noncyclic geometric gates implemented under (a-c) noncyclic evolution condition (i), under (d-f) condition (ii), and under (g-I) condition (iii), considering the combined impacts of universal systematic errors and residual crosstalk error.}
\label{Figure5}
\end{figure}

\section{Universally Error Robustness}
  
Achieving large-scale quantum computation necessitates simultaneously maintaining high gate fidelity and possessing strong error robustness. Consequently, a quantum gate's resilience against core error sources constitutes a critical metric for evaluating its performance. Next, focusing on key errors in quantum control, universal systematic errors and residual crosstalk error, we will identify the optimal noncyclic evolution condition that exhibits the strongest error robustness by evaluating their distinct impacts on geometric gate fidelity across different noncyclic evolution conditions.

The persistent presence of systematic errors remain an unavoidable practical challenge under current technological constraints. This limitation fundamentally stems from inherent defects in chip fabrication processes that are difficult to eradicate, coupled with inherent limitations in our knowledge of the controlled system model and its surrounding environment. Critically, systematic errors \cite{UCR1, UCR2} are not confined to errors in a single direction, such as driving amplitude error in the X-direction or detuning error in the Z-direction, but often arise from the combined effects of error sources in multiple directions. Here, we define this category of systematic errors, characterized by their multi-source and directional nature, as ``universal systematic error'', denoted in the form: \cite{ZZ1, ZZ2}
\begin{equation}
V_r=\lambda_m \vec{n} \cdot \vec{\sigma},
\end{equation}
%\lambda_m= \lambda \Omega_m
in which $\vec{n}$ is a randomly chosen unit vector, and $\lambda_m$ represents the error strength. Moreover, crosstalk error inevitably occurs when scaling qubits on multi-qubit chips due to unlocked inter-qubit interactions, leading to error accumulation and propagation. This presents a significant obstacle to current qubit integration efforts. A common experimental mitigation strategy involves operating qubits in a highly dispersive regime with large detuning. However, since this large detuning is approximate, the dynamics of adjacent qubits cannot be spatially factored. Consequently, persistent weak entanglement remains, manifesting as residual ZZ crosstalk \cite{ZZ1, ZZ2}. We consider a one-dimensional qubit chain, %\red{as shown in Fig. ??,} 
where an effective ZZ interaction of strength $\zeta_{zz}$ exists between target operating qubit $Q_1$ and adjacent spectator qubit $Q_{1a}$. This interaction is described by
\begin{equation}
V_{zz}=\frac{\zeta_{zz}}{2}\sigma_z\otimes\sigma^{a}_z,
\end{equation}
where $\sigma_z$ and $\sigma^{a}_z$ are the Pauli $Z$ operators of $Q_1$ and $Q_{1a}$, respectively.

Next, we will systematically compare the robustness of geometric quantum gates implemented under different noncyclic evolution conditions against universal systematic errors and residual crosstalk error. To quantitatively evaluate error sensitivity, the gate fidelity is defined as follows:
\begin{equation} 
F_e=|\textrm{Tr}(U^{\dagger} U_{e})| / |\textrm{Tr}(U^{\dagger} U)|,
\end{equation}
where $U$ represents the ideal evolution operator governed by the error-free Hamiltonian $\mathcal{H}(t)$, while $U_{e}$ corresponds to the error-affected evolution operator. Hereafter, we choose a simple $sin$-type pulse shape $\Omega(t)=\Omega_m \sin(\pi t/\tau)$ for convenience, where $\Omega_m$ represents the peak value of the pulse.

As analyzed in Section III regarding the construction of geometric gates under distinct noncyclic evolution conditions, it can be concluded that the geometric evolution trajectories corresponding to conditions (ii) and (iii) are uniquely determined. Crucially, however, the trajectory under noncyclic evolution condition (i) retains optimization flexibility: beyond parameters constrained for geometric gate construction, both the initial parameter $\chi_0$ and intermediate parameters $\{\chi_1,\chi_2\}$ can be optimized to mitigate error impacts on gate fidelity. For ease of analysis, we set the initial parameter $\chi_0=0$. Subsequently, based on the rotation angle $\theta$ of the target rotation gate, the optimization range for $\chi_1$ and $\chi_2$ can be set as $(0, \theta)$, while for the noncyclic geometric Hadamard gate, this range changes as $(0,\pi/2)$. Targeting the noncyclic geometric $H^i$ and $R^i_{x,y}(\pi)$ gates, Figure \ref{Figure4} identifies the intermediate trajectory parameters $\chi_1$ and $\chi_2$ that maximize gate robustness against universal systematic errors and residual crosstalk error. Through a balanced consideration of both types of errors, we can ultimately determine that $\{\chi_1, \chi_2\} = \{0.02\pi, 0.48\pi\}$ and $\{0.64\pi, 0.35\pi\}$ are respectively the optimal parameter settings for achieving robust geometric $H^i$ and $R^i_{x,y}(\pi)$ gates under the noncyclic evolution condition (i).

\begin{figure}[t]
  \centering
  \includegraphics[width=1.0\linewidth]{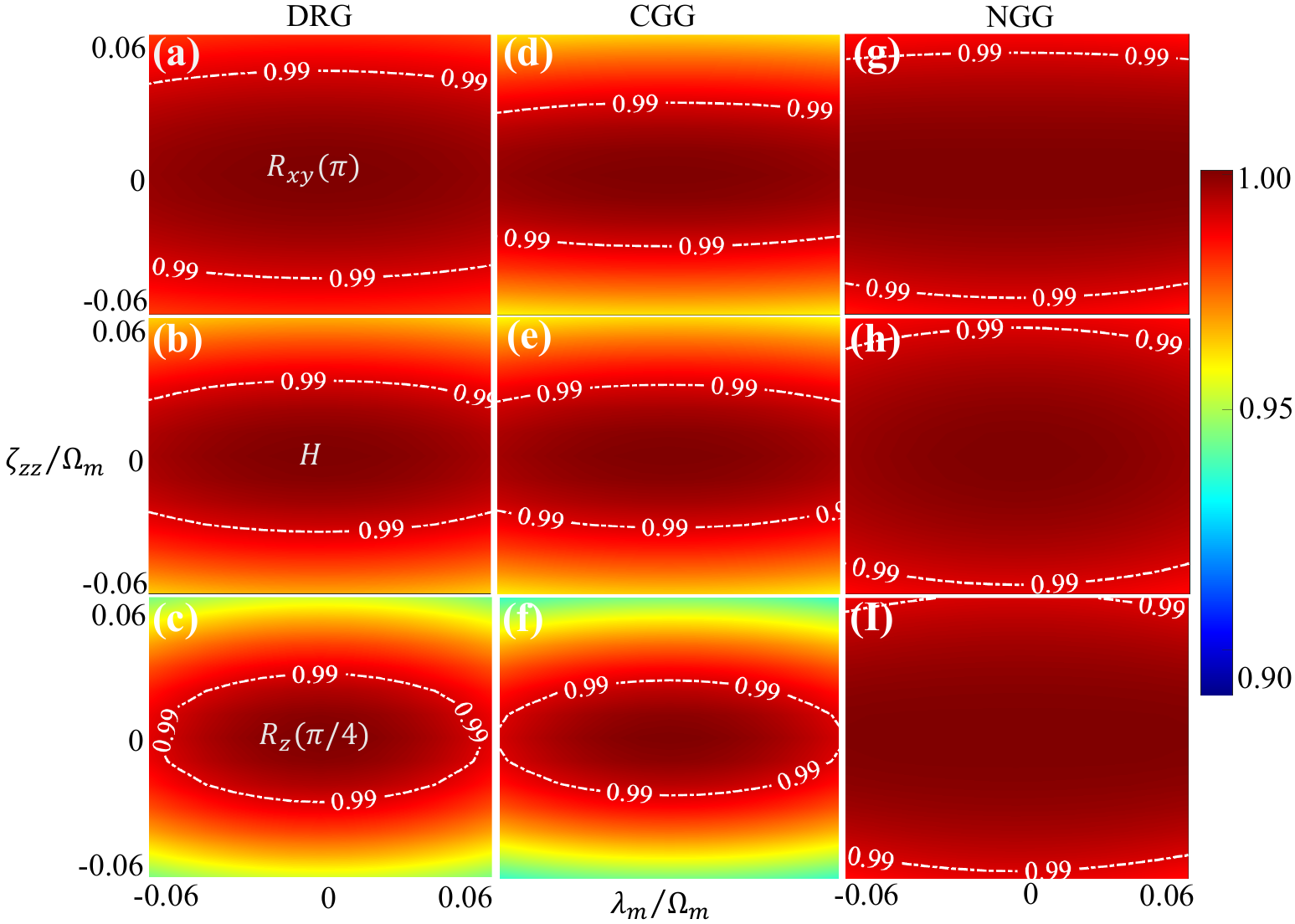}
  \caption{Comparison of gate robustness for (a-c) dynamical Rabi gates (DRG), (d-f) cyclic geometric gates (CGG), and (g-I) optimized noncyclic geometric gates (NGG), considering the combined impacts of universal systematic errors and residual crosstalk error.}
\label{Figure6}
\end{figure}

Therefore, we will comprehensively compare the resistance of noncyclic geometric gates implemented under three potential evolution conditions to universal systematic errors and residual crosstalk error, thereby screening for the condition that optimally enhances the gate robustness. As shown in Figs. \ref{Figure5}(a)-\ref{Figure5}(f), taking noncyclic geometric $H$ and $R_{x,y}(\pi)$ gates as examples, the gate robustness under different evolution conditions exhibits significant differences. Evidently, noncyclic geometric gates implemented based on condition (i) demonstrate stronger stability when simultaneously subjected to universal systematic errors and residual crosstalk error. This fully validates that systematically evaluating all potential noncyclic evolution conditions and conducting necessary screening on them under error influence is of critical importance for enhancing the error robustness of noncyclic geometric gates. In addition, as shown in in Figs. \ref{Figure6}(a)-\ref{Figure6}(I), compared to dynamical Rabi gates and conventional cyclic geometric gates (for details, see Appendix C), this enhanced robustness of noncyclic geometric gates maintains a significant advantage.

The rotation operation around the $Z$-axis can be substituted by a virtual $Z$ gate \cite{VZ}, a near-perfect $Z$ gate technique implemented by controlling the phase of the microwave drive used for $X$ and $Y$ rotations. Notably, considering the requirement for physical $Z$ gate in certain quantum algorithms and error-correction protocols, we also compare the error robustness of physical $Z$ gate under different schemes in Figs. \ref{Figure5}(g)-\ref{Figure5}(I) (using the $R_z(\pi/4)$ gate as an example). Specifically, the geometric $R_z(\pi/4)$ gate based on noncyclic evolution conditions (i)-(iii) can be implemented by setting the target gate parameter of the corresponding evolution operator and/or combining them. The results demonstrate that the noncyclic geometric $R^i_z(\pi/4)$ gate based on condition (i) still exhibits superior error resilience compared to noncyclic geometric gate based on other conditions, dynamical Rabi gate, and conventional cyclic geometric gate.

\section{High Gate Fidelity in Superconducting Implementation}
\begin{figure*}[t]
  \centering
  \includegraphics[width=0.85\linewidth]{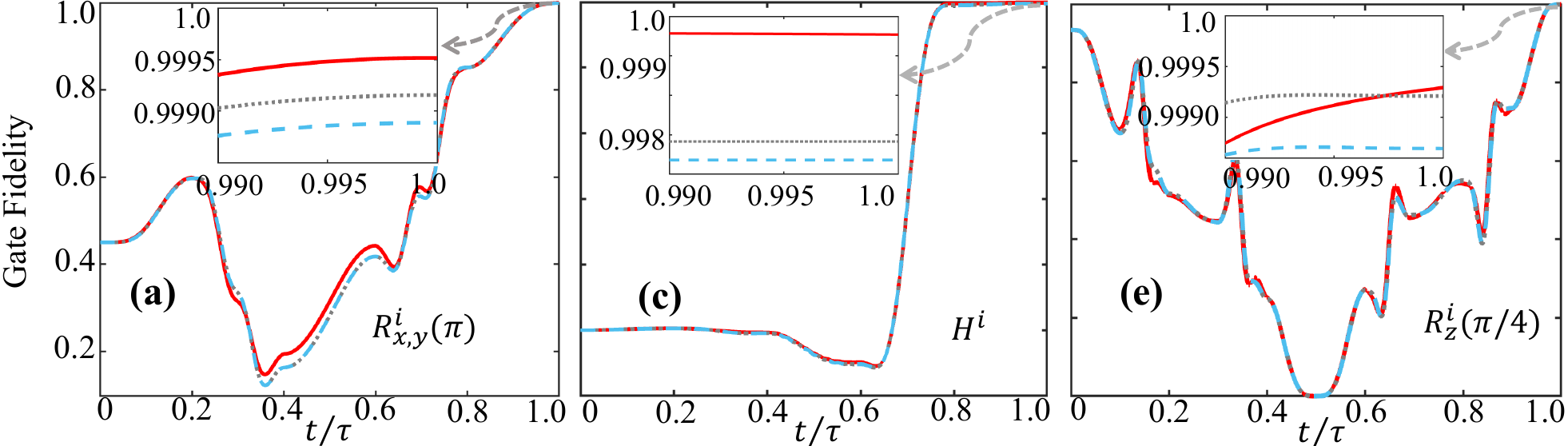}
  \caption{Gate infidelity of the implemented geometric (a) $R^i_{xy}(\pi)$, (b)$H^i$ and (c) $R_z(\pi/4)$, in which the gray dotted line considers only the leakage error, the blue dashed line accounts the decoherence effect and leakage error, without DRAG correction. Red solid line accounts these two effects with DRAG correction comprehensively.} %Subfigures (b) $R^i_{xy}(\pi)$, (d) $H^i$, and (f) $R_z(\pi/4)$ represent the robustness of noncyclic geometric quantum gates under the combined influence of universal system errors and residual crosstalk errors, respectively. }}
  \label{F8}
  \end{figure*}
  
In this section, we will further demonstrate the feasibility of high-fidelity physical realization of noncyclic geometric gates in superconducting quantum circuits. The impacts of the system's inherent leakage errors and decoherence effects on the construction of practical single- and two-qubit gates will be specifically considered. Firstly, the realization of the effective Hamiltonian $\mathcal{H}(t)$ in Eq. (\ref{GeneralH}) can be approximated by applying a microwave drive to a single superconducting transmon. The Hamiltonian parameters are thereby set to achieve the target noncyclic geometric evolution. Typically, the lowest two energy levels of a superconducting transmon constitute its qubit. However, the relatively weak anharmonicity between energy levels directly causes information leakage from the qubit subspace into higher-energy subspaces, with the leakage coupling between the $|1\rangle$ state and the non-computational basis state $|2\rangle$ being particularly severe. Meanwhile, the decoherence effects resulting from the inevitable interaction of the qubit with its environment during gate operations cannot be neglected. Therefore, we incorporate the system's inherent leakage errors and decoherence effects here by simulating the evolution using the master equation \cite{DRAG09}. The master equation we use is
\begin{align} \label{QME}
\begin{split}
\dot{\rho}_1&=-i \left[ \mathcal{H}(t) + \mathcal{H}_{\text{leak}}(t), \rho_1 \right] \\&+ \sum_{j=0}^{+\infty} \left\{ \frac{\kappa^1_-}{2} \mathcal{D}[\sqrt{j+1}|j\rangle\langle j+1|] + \frac{\kappa^1_z}{2} \mathcal{D}[j|j\rangle\langle j|] \right\},
\end{split}
\end{align}
where $\kappa^1_-$ and $\kappa^1_z$ denote the relaxation and dephasing rates of transmon qubit, respectively. $\rho_1$ is the density matrix, and $\mathcal{D}[\mathcal{A}] = 2\mathcal{A}\rho_1 \mathcal{A}^\dagger - \mathcal{A}^\dagger \mathcal{A} \rho_1 - \rho_1 \mathcal{A}^\dagger \mathcal{A}$ corresponds to the Lindblad superoperator associated with operator $\mathcal{A}$. The undesired leakage error arising from the coupling between the computational subspace and adjacent energy levels can be represented by the interaction term
\begin{align}
\begin{split}
\mathcal{H}_{\text{leak}}(t)
&=\sum_{j=2}^{+\infty}\left[(j-\frac{1}{2})\Delta(t)-\frac{j(j-1)\alpha}2\right]|j\rangle\langle j|  \\ &+\left[\frac12\Omega(t)e^{-i\phi(t)}\sum_{j=1}^{+\infty}\sqrt{j+1}|j\rangle\langle j+1|+\mathrm{H.c.}\right].
\end{split}
\end{align}

%\begin{figure}[b]
  %\centering
  %\includegraphics[width=1.0\linewidth]{F7.pdf}
  %\caption{\red{Performance of single-qubit noncyclic geometric gates. Gate fidelity results for $R^i_{x,y}(\pi)$, $H^i$, and $R^i_{z}(\pi/4)$, as a function of the tunable parameter $\Omega_m$ in subfigure (a)(b)(c), respectively.}}
  %\label{Fig.7}
%\end{figure}
Clearly, this coupling originates from insufficient system anharmonicity (failing to strictly satisfy $\alpha\gg\Omega(t)$). Furthermore, under finite coherence time constraints, it is impossible to circumvent the impact of leakage error solely by reducing the driving amplitude $\Omega(t)$. This is because excessively lowering the driving amplitude not only prolongs gate operation times, thereby exacerbating decoherence effects, but also increases the system's sensitivity to parameter fluctuations, ultimately increasing operational errors. This dilemma is mitigated by introducing the Derivative Removal via Adiabatic Gate (DRAG) technique \cite{DRAG09, DRAG2, DRAG3}, which suppresses such leakage errors by applying compensation to the original microwave pulses (for details, see Appendix D).
%As a result, while considering both leakage and decoherence effects, the infidelity of the quantum gate can be kept below the order of $10^{-3}$ after correction by DRAG method.

Building on state-of-the-art experimental fabrication capabilities, we adopt conservative parameter setting \cite{SQQC1, SQQC2, SQQC3}, setting the transmon anharmonicity to $\alpha=2\pi\times 320$ MHz. For simplicity, the relaxation rate and dephasing rate of the superconducting transmon qubit are assumed to be identical, i.e., $\kappa^1_{-}=\kappa^1_{z}=2\pi\times 2$ kHz. By numerically solving the quantum master equation in Eq. (\ref{QME}), the single-qubit final density matrix $\rho_{1}$, resulting from the target noncyclic geometric evolution, can be obtained. It is important to note that our numerical simulations are based on the full Hamiltonian $\mathcal{H}(t) + \mathcal{H}_{\text{leak}}(t)$, without employing any further approximations. 

Subsequently, we will utilize the resulting density matrix to systematically evaluate the fidelity of optimal noncyclic geometric gates under realistic conditions where leakage errors and decoherence effects are present. We consider a general initial state $|\psi_{i}\rangle=\cos \theta_1 |0\rangle+ \sin \theta_1 |1\rangle$, and the single-qubit-gate fidelity can be obtained by the calculation formula as $F_U=\frac{1}{2\pi}\int_{0}^{2\pi}\langle\psi_{f_U}|\rho_1|\psi_{f_U}\rangle d \theta_1$, in which $|\psi_{f_U}\rangle=U(\tau)|\psi_{i}\rangle $ represents the final state under the ideal noncyclic geometric gate operation with $U(\tau)=R^i_{x,y}(\pi)$, $H^i$, and $R^i_{z}(\pi/4)$. The above integration is numerically done for 1000 input initial states with $\theta_1$ being uniformly distributed within the range of $[0,2\pi]$. Under the parameter settings of $\Omega_m\approx2\pi\times 32$ MHz, $2\pi\times 25$ MHz, and $2\pi\times 32$ MHz, the noncyclic geometric $R^i_{x,y}(\pi)$, $H^i$, and $R^i_{z}(\pi/4)$ gates all achieve high gate fidelities exceeding 99.9\% as shown in Figs. \ref{F8}(a)-\ref{F8}(c). The comparison results in the insets respectively illustrate the relative contributions of leakage error sand decoherence effects to gate infidelity, as well as the correction effectiveness of DRAG technology. 
Building on the gate fidelities obtained by comprehensively considering leakage errors and decoherence effects in practical superconducting system  (for details, see Appendix E), %in Figs. \ref{F8}(a)-\ref{F8}(c), 
we further analyze the combined impacts of universal systematic errors $V_{r}$ and residual crosstalk error $V_{zz}$ within specific ranges. The results demonstrate that even considering these error factors, noncyclic geometric gates can still maintain remarkably high fidelities, exhibiting strong robustness against errors.

Without loss of generality, our scheme for implementing noncyclic geometric gates is applicable to realizing superconducting two-qubit entanglement gates, which, supplemented with single-qubit gates, allows for universal quantum computation. On the superconducting quantum processor, the interaction between two adjacent transmon qubits (denoted as $Q_1$ and $Q_2$) can be realized via simple capacitive coupling. To achieve parameter-tunable coupling between the qubits, an additional AC drive can be applied to transmon qubit $Q_1$. This is experimentally implemented by modulating the qubit via AC flux bias \cite{TunCoupleExp1, TunCoupleExp2}, enabling periodic modulation of its transition frequency in the form of $\omega_1(t)=\omega_1+\epsilon\sin(\nu+\varphi)$. It can be derived that the interaction Hamiltonian corresponding to the parameter-tunable coupling model takes the following form \cite{TunCoupleExp2}:
\begin{eqnarray} \label{EqHITwo}
\mathcal{H}_{I}(t)&&=\sum_{k=-\infty}^{+\infty} i^k J_k(\beta) \mathcal{G}_{12}\left\{|01\rangle\langle 10|e^{i \Delta_1 t} e^{i k(\nu t+\varphi)}\right.
\notag \\
&&\left.+\sqrt{2}|02\rangle\langle11|e^{i(\Delta_1-\alpha_2) t} e^{i k(\nu t+\varphi)}\right.
\notag \\
&&\left.+\sqrt{2}|11\rangle\langle20|e^{i(\Delta_1+\alpha_1) t} e^{i k(\nu t+\varphi)} \right\} +\mathrm{H.c.},
\end{eqnarray}
where $\Delta_1=\omega_2-\omega_1$ is the difference in the transition frequencies of qubits $Q_1$ and $Q_2$, $\omega_p$ and $\alpha_p$ are the transition frequency and anharmonicity of $Q_p$, respectively; $\mathcal{G}_{12}$ is the fixed coupling strength between the two transmon qubits. Furthermore, $J_k(\beta)$ denotes the Bessel function, with $\beta=\varepsilon/\nu$. It follows that since the dynamics within the two-qubit computational subspace $\{|00\rangle,|01\rangle,|10\rangle,|11\rangle\}$ involve interactions only within the single-excitation subspace $\{|01\rangle,|10\rangle\}$ and the two-excitation subspace $\{|02\rangle,|11\rangle,|20\rangle\}$, interactions in higher-excitation subspaces can be neglected. By selectively tuning the qubit drive frequency $\nu$, the above Hamiltonian can facilitate controllable effective two-level interactions either within the single-excitation subspace $\{|01\rangle,|10\rangle\}$ or within the two-excitation subspace $\{|02\rangle,|11\rangle\}$ (and $\{|11\rangle,|20\rangle\}$). Therefore, by adapting the parameter settings used to implement noncyclic geometric evolution (governed by the Hamiltonian $\mathcal{H}(t)$ in Eq. (\ref{GeneralH}) for single-qubit gates, we can correspondingly construct the noncyclic geometric $i$SWAP gate and controlled-phase gate.

For example, aiming to construct a noncyclic geometric $i$SWAP gate, we tune the qubit drive frequency $\nu$ to satisfy $\Delta_1-\nu=-\Delta_t$, with $|\Delta_t|\ll|\nu|$. By further applying appropriate rotating frame transformations and the rotating wave approximation (RWA), the target effective interaction Hamiltonian within the single-excitation subspace $\{|01\rangle,|10\rangle\}$ is obtained:
\begin{equation}
\mathcal{H}_{e}(t)=\frac{1}{2}\left(\begin{array}{cc}
	-\Delta_{e} & \Omega_{e} e^{-i \phi_{e}} \\
	\Omega_{e} e^{i \phi_{e}} & \Delta_{e}
\end{array}\right)
\begin{gathered}|01\rangle\\|10\rangle\end{gathered}
\end{equation}
where $\Omega_e=2J_1(\beta)\mathcal{G}_{12}$ and $\phi_{e}=(\Delta_t-\Delta_e)t+\varphi-\pi/2$ are effective coupling strength and phase. Under ideal conditions, the evolution operator arising from interactions within the two-excitation subspace $\{|02\rangle,|11\rangle,|20\rangle\}$ can be regarded as an identity operator. The controllable effective two-level Hamiltonian implemented in the single-excitation subspace $\{|01\rangle,|10\rangle\}$, as described above, can be adapted using the parameter settings used to implement the noncyclic geometric evolution (governed by Hamiltonian $\mathcal{H}(t)$ in Eq. (\ref{GeneralH}). This adaptation enables the realization of an effective rotation operation around the $X$-axis within this subspace. Consequently, the resulting evolution operator in the two-qubit computational subspace $\{|00\rangle,|01\rangle,|10\rangle,|11\rangle\}$ can be expressed as:
\begin{equation}
U_2 = 
\begin{pmatrix}
1 & 0 & 0 & 0 \\
0 & 0 & i & 0 \\
0 & i & 0 & 0 \\
0 & 0 & 0 & 1
\end{pmatrix},
\end{equation}
which corresponds to a nontrivial two-qubit $i$SWAP gate for quantum computation.

\begin{figure}[t]
\centering
\includegraphics[width=1.01\linewidth]{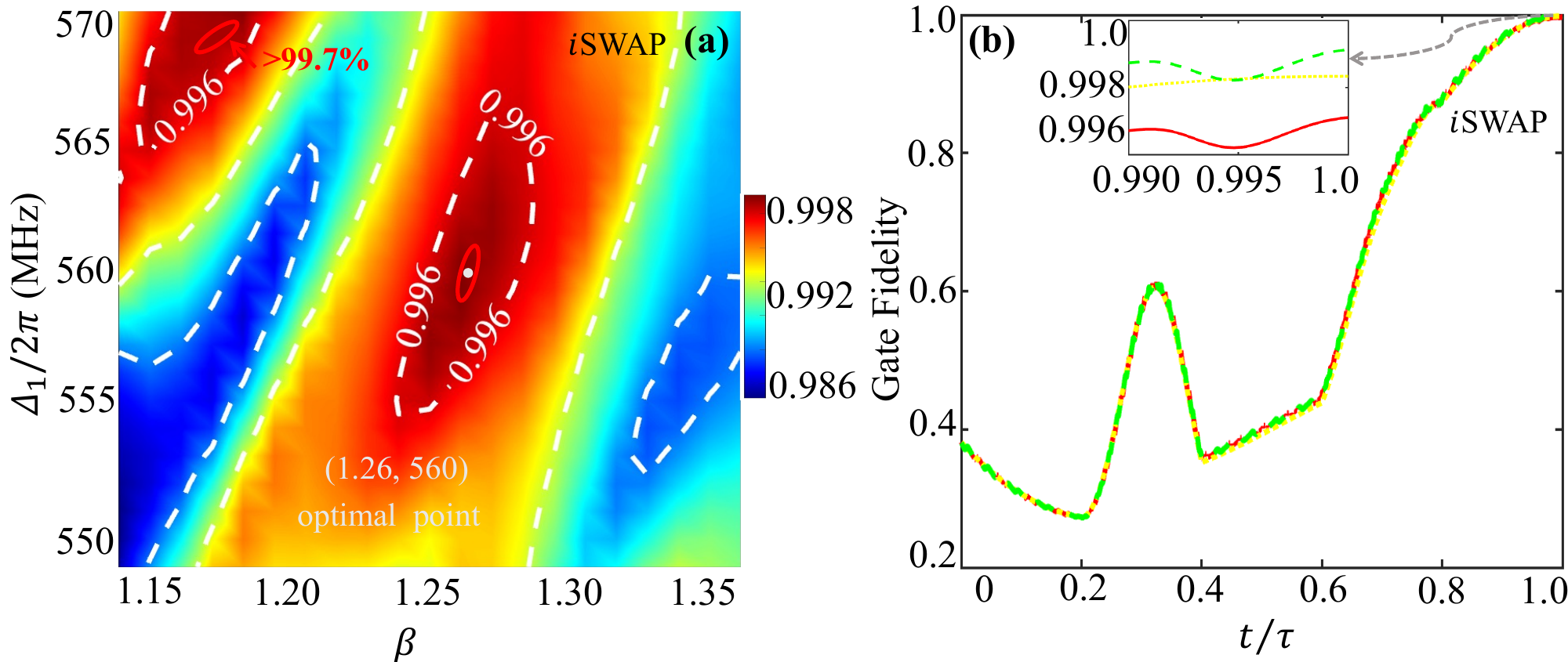}
\caption{(a) Gate fidelity of noncyclic geometric $i$SWAP gate in superconducting implementation as functions of tunable parameters $\left\{\Delta_1, \beta\right\}$. (b) Noncyclic geometric $i$SWAP gate under the optimal settings of $\Delta_1=2\pi\times 560$ MHz, $\beta\approx1.29$, where the red line, green line, and yellow line represent the results when considering both decoherence effects and leakage errors, leakage errors only, and decoherence effects only, respectively. }
 \label{F9}
\end{figure}

Similarly, we employ the quantum master equation \cite{DRAG09} to simulate the system evolution for evaluating the fidelity of the noncyclic geometric two-qubit gate in practical superconducting implementations, incorporating inherent leakage errors and decoherence effects of both transmons. In our simulation, the qubit parameters of transmon $Q_1$ remain consistent with the single-qubit case, while transmon $Q_2$ is configured with anharmonicity $\alpha_2=2\pi\times 300$ MHz and decoherence rates $\kappa_2^{-}=\kappa_2^{z}=2\pi\times 2$ kHz. Additionally, the inter-transmon coupling strength is set to $\mathcal{G}_{12}=2\pi\times 8$ MHz. Leakage errors primarily originate from undesired coupling interactions between the computational basis state $|11\rangle$ and the subspace ${|02\rangle, |20\rangle}$. As shown in Fig. \ref{F9}(a) by optimizing the drive frequency $\nu$ within a feasible range of the transition frequency difference between qubits $Q_1$ and $Q_2$, we identify a parameter region achieving fidelities exceeding 99.7\%. The comparative results in Fig. \ref{F9}(b) quantitatively show the relative contributions of leakage errors and decoherence effects to gate infidelity (exemplified at parameter point $\Delta_1=2\pi\times 560$ MHz, $\beta\approx1.29$). This confirms that our scheme for implementing noncyclic geometric gates can be effectively extended to practical superconducting systems for realizing high-fidelity two-qubit entanglement gates.

\section{Conclusion}
In summary, our work establishes noncyclic geometric gates as a promising approach for robust quantum computation, systematically overcoming the limitations of cyclic conditions and unlocking greater flexibility in evolution choice. By comprehensively evaluating all possible noncyclic evolution conditions, identifying optimal geometric trajectories, and rigorously quantifying resilience against universal systematic errors and residual crosstalk, we demonstrate that the optimized noncyclic geometric gates exhibit superior robustness compared to both dynamical Rabi gates and conventional cyclic geometric gates. Crucially, the high-fidelity physical implementation of these gates is validated in superconducting quantum circuits, specifically addressing the detrimental impacts of intrinsic leakage and decoherence. 

\bigskip

\acknowledgments

This work was supported by the National Natural Science Foundation of China (Grants No. 12305019, No. 12275090, and No. 12405009), Funding by Science and Technology Projects in Guangzhou (Grant No. 2024A04J4345), National College Students' Innovation Training Program (Grant No. 202410574037), and the Guangdong Provincial Quantum Science Strategic Initiative (Grant No. GDZX2203001).

\end{document}